\PassOptionsToPackage{svgnames}{xcolor}
\documentclass[11pt,reqno]{article}
\usepackage{amsmath,hyperref,amssymb, amsthm}
\usepackage{authblk}
\usepackage[all]{xy}
\usepackage{graphicx,url}
\usepackage{color}
\usepackage{slashed}
\usepackage{tikz}
\usepackage{lipsum}
\usepackage{caption}
\newcommand{\be}{\begin{eqnarray}}
\newcommand{\ee}{\end{eqnarray}}
\newcommand{\eeq}{\end{equation}}
\newcommand{\beq}{\begin{equation}}

\allowdisplaybreaks \numberwithin{equation}{section}

\DeclareSymbolFont{AMSa}{U}{msa}{m}{n}
\DeclareSymbolFont{AMSb}{U}{msb}{m}{n}
\DeclareMathSymbol{\fieldR}{\mathalpha}{AMSb}{"52}

\newcommand{\CA}{{\cal A}}

\newcommand{\CH}{\mathcal{H}}
\newcommand{\Hh}{\mathcal{H}}

\newcommand{\CI}{{\cal I}}
\newcommand{\CL}{{\cal L}}
\newcommand{\CN}{\mathcal{N}}

\newcommand{\calC}{\mathcal{C}}

\newcommand{\CW}{\mathcal{W}}

\newcommand{\zZ}{\mathsf{Z}}

\DeclareMathOperator{\Tr}{Tr}

\DeclareMathOperator{\Aut}{Aut}

\newcommand{\ZZ}{\mathbb{Z}}
\newcommand{\RR}{\mathbb{R}}
\newcommand{\CC}{\mathbb{C}}

\newcommand{\cTop}{\mathsf{Top}}

\def\beq{\begin{equation}}
\def\eeq{\end{equation}}
\def\bea{\begin{eqnarray}}
\def\eea{\end{eqnarray}}

\def\<{\langle}

\newtheorem{theorem}{Theorem}

\newtheorem{corollary}[theorem]{Corollary}

\DeclareMathOperator{\Hom}{Hom}
\DeclareMathOperator{\End}{End}
\DeclareMathOperator{\Rep}{Rep}

\usepackage{tcolorbox}
\tcbuselibrary{skins,breakable}
\usetikzlibrary{shadings,shadows}

    {\endtcolorbox}

    {\endtcolorbox}

    {\endtcolorbox}

\title{Vertex algebras, topological defects, and Moonshine}
\author{Roberto Volpato\thanks{volpato@pd.infn.it}}
\affil{\small Dipartimento di Fisica e Astronomia `Galileo Galilei', Universit\`a di Padova \& INFN, sez. di Padova, Via Marzolo 8, 35131, Padova, Italy}

\begin{document}

\maketitle

\begin{abstract}
	We discuss topological defect lines in holomorphic vertex operators algebras and superalgebras, in particular Frenkel-Lepowsky-Meurman Monster VOA $V^\natural$ with central charge $c=24$, and Conway module SVOA $V^{f\natural}$ with $c=12$. First, we consider duality defects in $V^\natural$ for all non-anomalous  Fricke elements of the Monster group, and provide a general formula for the corresponding defect McKay-Thompson series. Furthermore, we describe some general properties of the category of defect lines preserving the $N=1$ superVirasoro algebra in $V^{f\natural}$. We argue that, under some mild assumptions, every such defect in $V^{f\natural}$ is associated with a $\ZZ$-linear map form the Leech lattice to itself. This correspondence establishes a surjective (not injective) ring homomorphism between the Grothendieck ring of the category of topological defects and the ring of Leech lattice endomorphisms. Finally, we speculate about possible generalization of the Moonshine conjectures that include topological defect lines.
\end{abstract}


\section{Introduction}

In recent years, various generalizations of the idea of global symmetry in Quantum Field Theory have been proposed \cite{Gaiotto:2014kfa} (see the recent reviews \cite{McGreevy:2022oyu,Cordova:2022ruw,Schafer-Nameki:2023jdn,Brennan:2023mmt,Bhardwaj:2023kri,Shao:2023gho} for further information and references). In the context of two dimensional conformal field theory (CFT), the most general concept of symmetries are represented by topological line defects (often referred to as non-invertible symmetries or categorical symmetries). As we will review in section \ref{s:TDLreview}, the latter can be thought of as objects in a fusion category; the case of an ordinary (finite) group $G$ of invertible symmetries is recovered as a particular case of the category ${\mathrm Vec}_G$ of $G$-graded finite dimensional complex vector spaces. The study of defects in 2D CFT has actually a much longer history than their analogues in higher dimensional quantum field theories. Some examples were already discussed in \cite{Verlinde:1988sn} and a systematic treatment was proposed in \cite{Petkova:2000ip,Frohlich:2004ef,Frohlich:2006ch,Frohlich:2009gb}; more recent results and discussions include \cite{Chang_2019,Bhardwaj:2017xup}.

Vertex operators algebras (VOA) represent one of the main proposals for a rigorous mathematical definition of two dimensional conformal field theory. Thus, it is natural to expect the concept of  topological defect in physics to correspond to a generalization of the idea of automorphism of a VOA -- see for example \cite{Moller:2024xtt} for a recent discussion. One of the most obvious applications of these `generalized symmetries' in VOA is in the context of the Moonshine conjectures \cite{ConwayNorton1979}. Surprisingly, this idea has not been considered until very recently \cite{Lin:2019hks,Fosbinder-Elkins:2024hff}. 

The goal of this article is to discuss some recent applications of the idea of topological defects to holomorphic vertex operators algebras and superalgebras, and in particular to possible generalization of the Moonshine conjectures. We also describe some of the open questions in this area and make various speculations inspired by physics intuition.

The structure of the article is as follows:
\begin{itemize}
	\item In section \ref{s:TDLreview}, we review  the main properties of topological defects in two-dimensional conformal field theories, and discuss how these could be formalized in the context of holomorphic vertex operator algebras.
	\item In section \ref{s:Monstrous} we consider the so called `duality defects' in Frenkel-Lepowsky-Meurman Monster module $V^\natural$. On general grounds, for every non-anomalous Fricke element $g$ of the Monster group $\mathbb{M}$, one expects a duality defect $\CN_g$ such that the $\CN_g$ and $\CL_{g^k}$ generate a Tambara-Yamagami category \cite{Tambara:1998} for the group $\langle g\rangle\cong\ZZ_N$. We derive an explicit general formula for the McKay-Thompson series $T_{\CN_g}(\tau)$ for all such duality defects,  generalizing previous results in \cite{Lin:2019hks} and \cite{Fosbinder-Elkins:2024hff}.
	\item  In section \ref{s:Moonshine} we speculate about possible extensions of the Monstrous Moonshine conjectures that include topological defects in $V^\natural$.
	\item In section \ref{s:Conway}, we present some new results about topological defects preserving the $N=1$ superVirasoro algebra in the Conway module $V^{f\natural}$, the only holomorphic vertex operator superalgebra of central charge $c=12$ with no operators of conformal weight $1/2$. The proofs of these results will appear elsewhere \cite{Angius:2024xxx}. 
\end{itemize}

\section{Generalities on topological defect lines in 2D CFT}\label{s:TDLreview}

In this section, we review some properties of topological line defects in two dimensional CFT, both from a physics and a mathematical perspective. See for example \cite{Chang_2019} and \cite{Moller:2024xtt} for more details.

In this article, we consider unitary two-dimensional conformal field theories at central charges $(c,\tilde c)$ with a unique vacuum, defined on a (Euclidean) worldsheet $\Sigma$, which is a Riemann surface. In fact, we will further restrict to worldsheets $\Sigma$ with the topology of a cylinder $S^1\times\RR$, a (Riemann) sphere $\hat\CC$, or a torus $S^1\times S^1$. In this section, we assume that the CFT is bosonic; the generalization to CFTs with fermions is discussed in section \ref{s:Conway}.

The most general correlation functions in these CFTs contain both local operators $O(z)$ with support on a point $z\in\Sigma$ and defects $\CL(\gamma)$ supported on oriented lines $\gamma\subset \Sigma$. The line $\gamma$ might be either closed or open; in the latter case, some suitable `defect starting' and `defect ending' operators should be inserted at the endpoints. A line defect $\CL$ is topological if all correlation functions with an insertion $\CL(\gamma)$ are invariant under continuous deformations of the support line $\gamma$, provided that no other defect or point-like operator  is crossed in the deformation. More generally, one can allow for $k$-junction operators ($k=1,2,3,\ldots$), i.e. point-like operators attached to $k$ topological defect lines, and consider correlation functions with the insertion of networks of defects.

The holomorphic and anti-holomorphic stress-energy tensor operators $T(z)$ and $\tilde T(\bar z)$ are always `transparent' to any topological defect $\CL$, in the sense that a correlation function does not change when $\CL$ is moved across the support of a $T$ or $\tilde T$ insertion. For a given defect $\CL$, it might happen that there are other holomorphic and/or antiholomorphic fields $\phi(z)$ or $\tilde \phi(\bar z)$, besides $T(z)$ and $\tilde T(z)$, that are `transparent' with respect to $\CL$. The algebras $\CA$ and $\tilde \CA$ generated, respectively, by the modes $\phi_n$ and $\tilde \phi_n$ of such fields are called the \emph{preserved} (chiral and antichiral) algebras of $\CL$.  It follows that both $\CA$ and $\tilde \CA$ always contain a copy of the Virasoro algebra $Vir_c$ and $Vir_{\tilde c}$ at central charges $c$ and $\tilde c$.

\begin{figure}[h]\includegraphics[width=.7\textwidth]{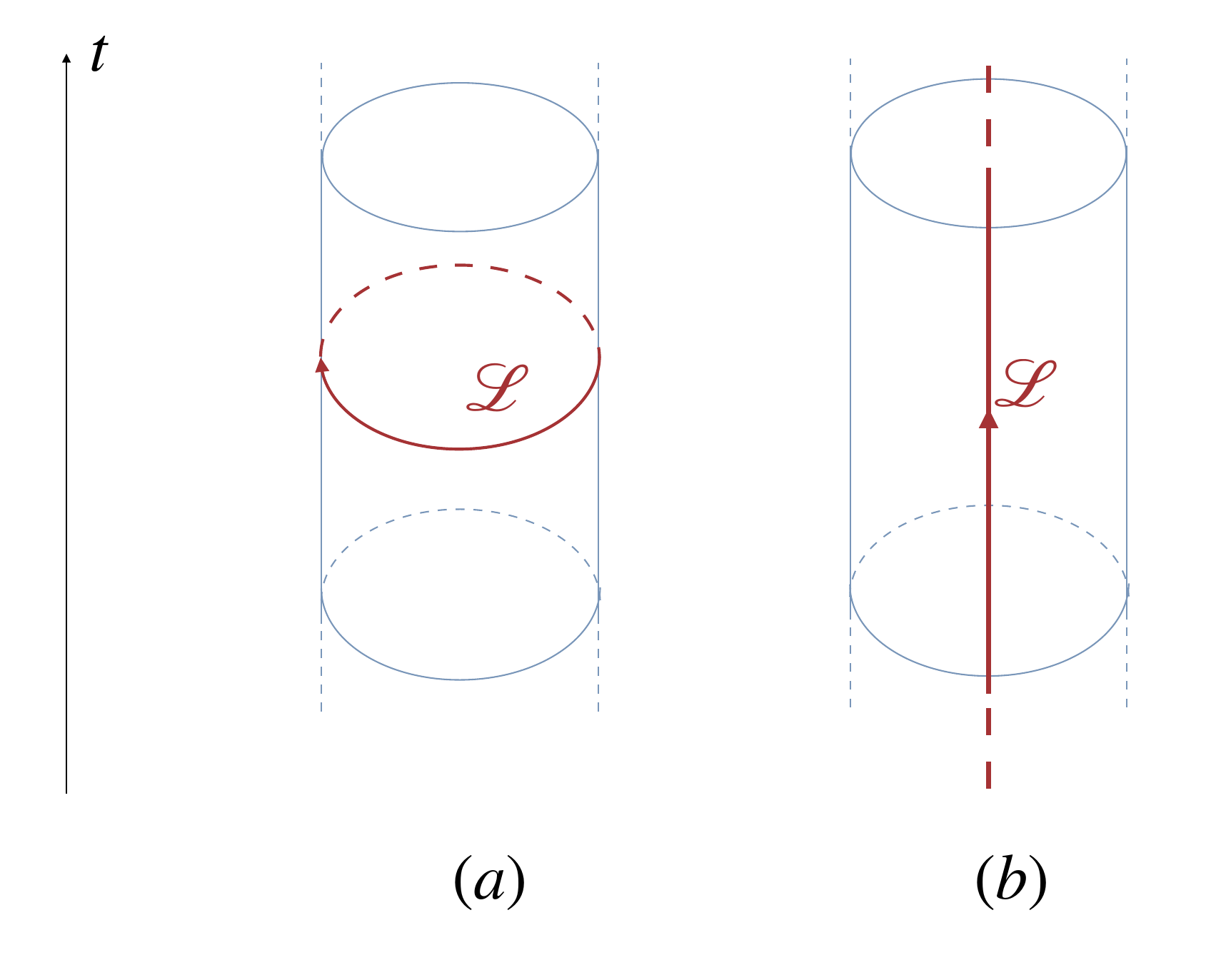}
	\caption{A topological defect line $\CL$ inserted in a worldsheet with the topology of a cylinder $S^1\times \RR$. (a) When the defect line is  wrapped once around the circle $S^1$ at fixed time $t$, it corresponds to the insertion of a linear operator $\hat\CL:\CH\to \CH$ in the time ordered correlation function. (b) When the line extends along the time direction, it corresponds to replacing the ordinary Hilbert space of states $\CH$ with a $\CL$-twisted space $\Hh_\CL$.}\label{fig:01}\end{figure}

Consider the CFT defined on a cylinder $S^1\times \RR$, with the $\RR$ factor interpreted as the Euclidean time direction, and let $\CH$ be the Hilbert space of states on the circle $S^1$ (a space slice at fixed time). By the state-operator correspondence $|\Phi\rangle \leftrightarrow \Phi(z,\bar z)$, $\CH$ can also be interpreted as the space of local operators.  The insertion of a topological defect $\CL$ line winding once the circle $S^1$ at some fixed time $t$ corresponds to the insertion of a linear operator $\hat\CL:\CH\to \CH$ in the corresponding time-ordered correlation function (see figure \ref{fig:01} (a)).
Thus, with each topological defect $\CL$ is associated a linear map $\hat\CL:\CH\to \CH$ that commutes with the preserved algebras $\CA$ and $\tilde \CA$. In particular, the vacuum $|0\rangle\in \CH$ must be an eigenstate for $\hat\CL$
\be \hat\CL|0\rangle=\langle \CL\rangle |0\rangle\ .
\ee With our assumptions of unitarity and uniqueness of the vacuum, the eigenvalue $\langle \CL\rangle$ (quantum dimension) is always real and satisfies $\langle \CL\rangle\ge 1$ \cite{Chang_2019}. 

One can also insert a defect $\CL$ on an infinite line extending along the time direction (see figure \ref{fig:01} (b)). This insertion correspond to modifying the space of states on $S^1$, which becomes a `$\CL$-twisted' space $\Hh_{\CL}$. By the state-operator correspondence, $\Hh_{\CL}$ is a also the space of point-like operators starting a defect $\CL$ ($1$-junction operators). For each $\CL$, the space $\Hh_{\CL}$ is a module for the algebras $\CA$ and $\tilde \CA$ preserved by $\CL$.

In every CFT, there is always an identity defect $\CI$, that can be defined by the property that its insertion does not modify any correlation function. It corresponds to the identity operator $\hat\CI:\Hh\to \Hh$ and the space $\CH_{\CI}$ coincides with $\CH$. 

\begin{figure}[h]\includegraphics[width=.9\textwidth]{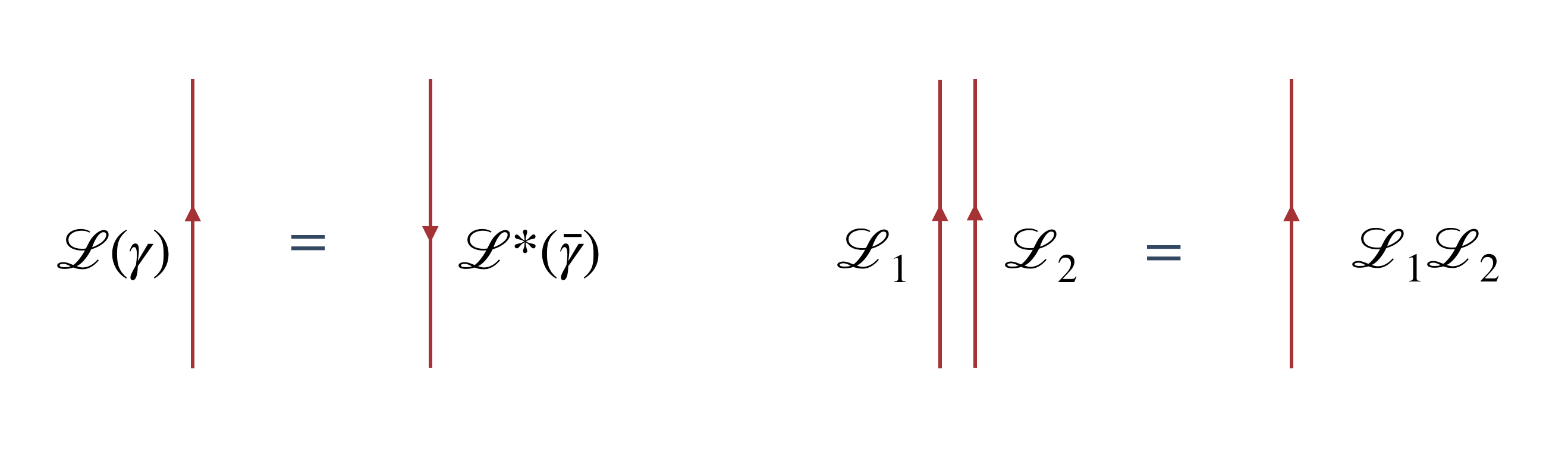}
	\caption{On the left, inserting a defect $\CL$ along a line $\gamma$ is the same as inserting the dual defect $\CL^*$ along the line $\bar\gamma$ with reversed orientation. On the right, the fusion of two topological line defects $\CL_1$ and $\CL_2$ into a single one $\CL_1\CL_2$.}\label{fig:02}\end{figure}

From an abstract point of view, the different kinds of topological defects $\CL$ preserving a given chiral and antichiral algebras $\CA$ and $\tilde \CA$, can be represented as objects in a fusion category $\calC$. We assume that $\calC$ is a spherical semisimple fusion category. In this article, we only consider unitary CFTs; in this case, we also assume that the fusion category is unitary.\footnote{A fusion category is unitary if all the complex vector spaces of morphisms are endowed with a positive definite hermitian form, and all fusion matrices are unitary.} For each object (defect) $\CL$ in this category, there is a dual defect $\CL^*$, such that in any correlation function the insertion of $\CL(\gamma)$ on the oriented line $\gamma$ is equivalent to the insertion of $\CL^*$ on the line $\bar\gamma$ with opposite orientation (see figure \ref{fig:02} left). The operator $\hat\CL^*$ is the adjoint of $\hat\CL$, and the $\CL^*$-twisted sector $\CH_{\CL^*}$ is the dual space of $\CH_{\CL}$.

The fusion product $\CL_1\CL_2$ of any two defects $\CL_1$ and $\CL_2$ is the defect obtained by moving two parallel defects $\CL_1$ and $\CL_2$ very close to each other  (see figure \ref{fig:02} right). The operator $\widehat{\CL_1\CL_2}$ is just the product $\hat\CL_1\hat\CL_2$ of linear operators, and the $\CL_1\CL_2$-twisted space $\CH_{\CL_1\CL_2}$ is isomorphic to (a suitable version of) tensor product
\be \CH_{\CL_1\CL_2}\cong \CH_{\CL_1}\otimes \CH_{\CL_2}\ .
\ee The identity defect $\CI$ is the unit of the fusion product. In the tensor category literature, the fusion product is usually denoted as $\CL_1\otimes\CL_2$, while in physics the symbol $\otimes$ is usually omitted, as in $\CL_1\CL_2$; we will use both notations, depending on the context. For any three defects $\CL_i$, $\CL_j$, $\CL_k$, the associator 
\beq \alpha_{i,j,k}:(\CH_{\CL_i}\otimes \CH_{\CL_j})\otimes \CH_{\CL_k}\stackrel{\cong}{\longrightarrow}\CH_{\CL_i}\otimes (\CH_{\CL_j}\otimes \CH_{\CL_k})\ ,
\eeq is possibly non-trivial, and satisfies a pentagonal identity.  The space $\CH_{\CL_1\CL_2\cdots\CL_k}$ is interpreted as the space of $k$-junction operators attached to $k$ outgoing lines $\CL_1,\ldots, \CL_k$. We are particularly interested in the subspace of \emph{topological} $k$-junction operators, i.e. operators in $\CH_{\CL_1\CL_2\cdots\CL_k}$ with vanishing conformal weights ($L_0$, $\bar L_0$-eigenvalues). From the perspective of tensor categories, the subspace of topological junctions in $\CH_{\CL_1\CL_2\cdots\CL_k}$ is identified with the $\CC$-vector space of morphisms $\Hom(\CI, \CL_1\otimes\CL_2\otimes\cdots\otimes\CL_k)$, i.e.
\be \Hom(\CI, \CL_1\otimes\CL_2\otimes\cdots\otimes\CL_k)=\{v\in \CH_{\CL_1\cdots\CL_k}\mid L_0v=0=\bar L_0v\}\ .
\ee 
\begin{figure}[h]\includegraphics[width=.7\textwidth]{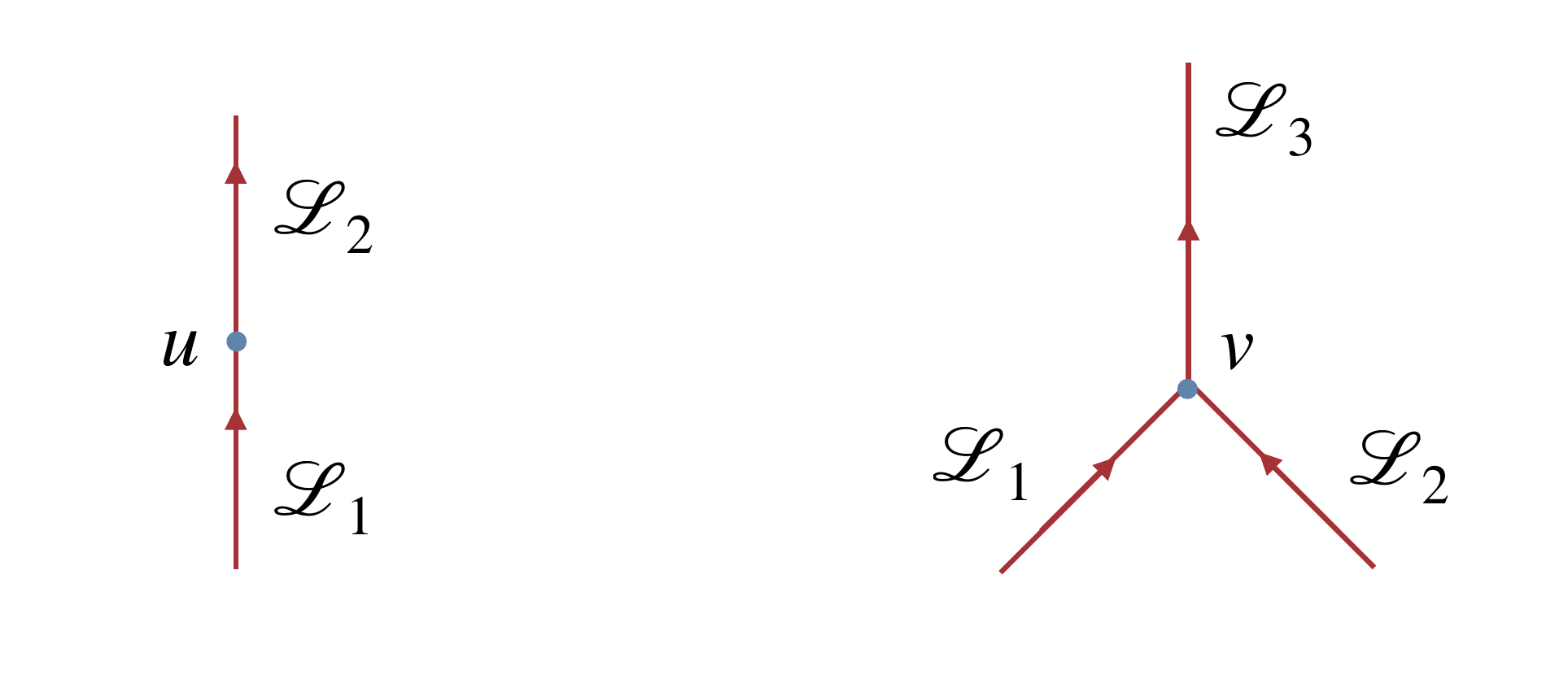}
	\caption{On the left, a topological $2$-junction $u\in \Hom(\CL_1,\CL_2)$ corresponds to a linear map $u:\CH_{\CL_1}\to \CH_{\CL_2}$ between twisted spaces. On the right, a topological $3$-junction $v\in \Hom (\CL_1\otimes \CL_2,\CL_3)$ corresponds to a linear map $v:\CH_{\CL_1}\otimes \CH_{\CL_2}\to \CH_{\CL_3}$.}\label{fig:03}\end{figure}
In particular, given two objects $\CL_i$ and $\CL_j$, the space $\Hom (\CL_i,\CL_j)$ of morphisms between $\CL_i$ and $\CL_j$  is identified with the space of topological $2$-junctions
\be \Hom (\CL_i,\CL_j)\cong \Hom (\CI,\CL_j\otimes \CL_i^*)\subset \CH_{\CL_j\CL_i^*}\ ,\ee whose elements corresponds to morphisms $\CH_{\CL_i}\to \CH_{\CL_j}$ of twisted modules for the algebra of local operators. More generally, we have the identification
\be \Hom (\CL_i\otimes \CL_j,\CL_k)\cong \Hom (\CI,\CL_k\otimes (\CL_i\otimes \CL_j)^*)\subset \CH_{\CL_k(\CL_i\CL_j)^*}\ ,
\ee between morphisms $\CL_j\otimes\CL_j\to \CL_k$ and $3$-junction topological operators.
For each $\CL$, one has
\be \dim \Hom(\CI,\CL\otimes\CL^*)=\dim\Hom (\CL,\CL) \ge 1
\ee
because of the identity morphism $\CH_{\CL}\to \CH_{\CL}$. A defect $\CL$ is called \emph{simple} if $\Hom(\CI,\CL\otimes\CL^*)\cong \CC$.

We can also define a superposition $\CL_1+\CL_2$ of two defects $\CL_1$ and $\CL_2$ such that 
\be\widehat{\CL_1+\CL_2}=\hat\CL_1+\hat\CL_2\ ,\qquad \CH_{\CL_1+\CL_2}=\CH_{\CL_1}\oplus \CH_{\CL_2}\ .\ee
We assume that our category is semisimple, i.e. that every defect $\CL$ can be decomposed into a (finite) superposition of simple defects. A defect is simple if and only if it is indecomposable, and for any two simple defects $\CL$ and $\CL'$, one has 
\be \dim\Hom(\CL,\CL')=\begin{cases}
	1 &\text{if } \CL\cong \CL'\ ,\\
	0 &\text{otherwise}\ .
\end{cases}
\ee Furthermore, given two simple defects $\CL_i$ and $\CL_j$, one can decompose their fusion product $\CL_i\CL_j$ into simple defects as
\beq \CL_i\CL_j=\sum_{\text{simple }\CL_k} N_{ij}^k \CL_k\ ,
\eeq where the fusion coefficients $N_{ij}^k\in \ZZ_{\ge 0}$ are given by the dimensions of the spaces of topological $3$-junction operators
\be N_{ij}^k=\dim \Hom(\CL_i\otimes\CL_j,\CL_k)=\dim \Hom(\CI,\CL_k\otimes(\CL_i\otimes\CL_j)^*)\ .
\ee For every (not necessarily simple) $\CL$, one has $\Hom(\CL,\CL)=\Hom(\CL\otimes\CL^*,\CI)$, so that the fusion between $\CL$ and $\CL^*$ always contains the identity defect with multiplicity $n:=\dim\Hom(\CL,\CL)\ge 1$
\be \CL\CL^*=n\CI+\ldots\ ,
\ee and $n=1$ if and only if $\CL$ is simple.  Quantum dimensions provide a $1$-dimensional representation of the fusion ring
\be \langle \CL_i\rangle\langle \CL_j\rangle=\sum_{\text{simple }\CL_k} N_{ij}^k\langle \CL_k\rangle\ .
\ee
Usually, in the definition of a fusion category one requires the number of isomorphism classes of simple objects to be finite. In physics, and in particular for topological defects in non-rational CFTs, one often needs to drop this condition. We will anyway use the term fusion category in these cases as well. We will in all cases require that the quantum dimension $\langle \CL\rangle$ is finite. By unitarity and uniqueness of the vacuum, one has $\langle \CL\rangle \ge 1$, and this implies that only a finite number of simple defects appear in the decomposition of any fusion product $\CL_i\CL_j$.

A topological defect is \emph{invertible} if it is simple and if $\CL\CL^*=\CI$. This implies (and, in fact, is equivalent) that the corresponding linear operator $\hat\CL$ is invertible. Furthermore, a topological defect is invertible if and only if its quantum dimension is $1$.

A simple example of fusion category of topological defects is the one associated with a finite group $G$ of global symmetries of the CFT. Simple defects $\CL_g$ are in one-to-one correspondence with  elements $g\in G$. The fusion product is given by the group law
\be \CL_g\CL_h=\CL_{gh}\ ,
\ee and the dual is $\CL_g^*=\CL_{g^{-1}}$. This implies that all $\CL_g$ are invertible and therefore have quantum dimension $\langle \CL_g\rangle=1$. Vice versa, if all simple defects in a fusion category are invertible, then the defects are associated with a group of symmetries $G$. The operators $\hat\CL_g:\CH\to \CH$ are just the linear operators representing the action of the symmetry group $G$ on the space of states $\CH$, and $\CH_{\CL_g}\equiv \CH_g$ are the $g$-twisted sectors. For any three simple $\CL_g,\CL_h,\CL_k$, the associator is given by multiplication by $\alpha(g,h,k)\in U(1)$. The pentagon identity implies that the function $\alpha:G\times G\times G\to U(1)$ is a $3$-cocycle representing a class $[\alpha]\in H^3(G,U(1))$. In physics language, a non-trivial class $[\alpha]$ is called a 't Hooft anomaly, and represents an obstruction to gauging the CFT (i.e. taking an orbifold) by the group $G$.  The  fusion category generated by these defects is therefore $\mathrm{Vec}_G^\alpha$, the fusion category of $G$-graded vector spaces with non-trivial associator $\alpha$.

\bigskip

Let us now discuss how this construction specializes to the case of a \emph{chiral} CFT, generated only by holomorphic fields. In this case, the CFT can be described in terms of a holomorphic VOA $V\equiv \CH$, i.e. a strongly rational  (rational, $C_2$-cofinite and of CFT type) VOA with only one admissible irreducible module, isomorphic to $V$ itself. Our goal is to provide a precise description the fusion category $\calC_W(V)$ of topological defects in $V$ preserving a conformally embedded subVOA $W\subset V$. We will be able to do so under the assumption that the subVOA $W$ is also strongly rational, so that the category $\Rep(W)$ of $W$-modules is a modular tensor category. Then, one can regard $V$ itself as a $W$-module, corresponding to an object $V\in \Rep(W)$. This object is enriched with a multiplication morphism $m:V\otimes V\to V$ given by the VOA strucure, as well as a unit morphism $\eta:W\to V$, so that $V$ can be interpreted as a symmetric haploid algebra object in $\Rep(W)$. For each defect $\CL\in \calC_W(V)$, the $\CL$-twisted space $V_\CL$ is also a $W$-module, and the $3$-junction operators $l_{\CL}\in \Hom(\CI\otimes \CL,\CL)$ and $r_{\CL}\in \Hom(\CL\otimes \CI,\CL)$ are associated with morphisms $l_{\CL}: V\otimes V_\CL\to V_{\CL}$ and $r_{\CL}: V_\CL\otimes V\to V_{\CL}$ in the category $\Rep(W)$. This structure turns each $V_\CL\in \Rep(W)$ into a $(V,V)$-bimodule for the algebra object $V$. The category $\mathrm{BiMod}_W(V)$ of such $(V,V)$-bimodules has naturally the structure of a fusion category. This arguments suggest that there is an equivalence of fusion categories \be \calC_W(V)\cong \mathrm{BiMod}_W(V)\ ,\ee and in fact one can use this equivalence to provide a more precise definition of the category $\calC_W(V)$. We stress that the construction of $\calC_W(V)$ we just described is not new, and is in fact a reformulation of a circle of ideas that appeared in several different versions in the physics literature. See \cite{Frohlich:2004ef,Frohlich:2006ch,Fuchs:2002cm,Fuchs:2003id,Fuchs:2003id,Fuchs:2004dz,Fuchs:2004xi,Bhardwaj:2017xup,Diatlyk:2023fwf,Choi:2023xjw,Rayhaun:2023pgc,Moller:2024xtt} for more discussions and references.

In the particular case where $W$ is the fixed point subalgebra $W=V^G\subseteq V$ with respect to a (solvable) finite group of automorphisms $G\subseteq \Aut(V)$, all simple objects in  $\calC_W(V)$ are invertible and can be identified with the $g$-twisted modules $V_g$, $g\in G$. In particular, $\calC_W(V)$ is isomorphic to the category $\mathsf{Vec}_G^\alpha$ of $G$-graded finite dimensional vector spaces, twisted by a suitable $3$-cocycle $\alpha$ defining a class $[\alpha]\in H^3(G,U(1))$.  It would be desirable to have a concrete description of the $(V,V)$-bimodule $V_\CL$ as a `$\CL$-twisted module for $V$',  even in the general case where $\calC_W(V)$ contains some non-invertible simple defect $\CL$. Based on physical intuition, we expect such a description to include a proper definition of `$z$-dependent $V_\CL$-endomorphisms' $Y^\CL(v,z):V_\CL\to V_\CL$ for all $v\in V$, satisfying a suitable set of axioms, analogous to the concept of $g$-twisted vertex operators for invertible $\CL_g$. To the best of my knowledge, however, a rigorous definition of such $\CL$-twisted vertex operators $Y^\CL$ in the context of vertex operator algebras has not been proposed so far.

In the next sections, we will also consider categories $\calC_W(V)$ of topological defects whose preserved subVOA $W\subset V$ is not rational. In this case, the definition above does not apply directly. We still expect $\calC_W(V)$ to be a unitary tensor category, but possibly with an infinite number of isomorphism classes of simple objects.

\section{Duality defects in the Monster module}\label{s:Monstrous}

In this section, we discuss some exaples of (non-invertible) topological defect lines in Frenkel-Lepowsky-Meurman Monster module $V^\natural$ \cite{FLM1988}. The latter is a holomorphic VOA with central charge $c=24$ without any operators of conformal weight $1$ (currents) -- conjecturally, it is the only VOA with these properties. Its automorphism group $\Aut(V^\natural)$ is a finite group isomorphic to the Monster group $\Aut(V^\natural)\cong \mathbb{M}$. For each $g\in \Aut(V^\natural)$, the McKay-Thompson series $T_g(\tau)$ is given by\footnote{Historically, the McKay-Thompson series and the Monstrous moonshine conjecture were introduced before the FLM module $V^\natural$ was discovered.}
\beq T_g(\tau):=\Tr_{V^\natural}(q^{L_0-1}g)\ .
\eeq The Monstrous moonshine conjecture \cite{ConwayNorton1979} (now a theorem proved by Borcherds \cite{Borcherds1992}) states that all such McKay-Thompson series are modular functions for a group $\Gamma_g\subset SL_2(\RR)$ commensurable with $SL_2(\ZZ)$ and that is genus zero, i.e. such that the quotient $\mathbb{H}/\Gamma_g$ of the upper-half plane has the topology of a sphere (with punctures). Furthermore, $T_g$ are actually Hauptmodul for the corresponding $\Gamma_g$, i.e. they generate the field of meromorphic functions on $\mathbb{H}/\Gamma_g$. 

In CFT language, the McKay-Thompson series $T_g$ can be interpreted as the partition function of the theory $V^\natural$ when the CFT is defined a torus $\mathbb{T}=S^1\times S^1$ with the insertion of the invertible topological  defect $\CL_g$ inserted along a `space-circle' $S^1$ at a fixed Euclidean time. More generally, for any commuting pair $g,h\in \Aut(V^\natural)$, one can define a $g$-twisted $h$-twining partition function on the torus (a generalized McKay-Thompson series) $T_{g,h}$, corresponding to inserting a defect $\CL_h$ along the space-like circle and a defect $\CL_g$ along the Euclidean time circle, and given by
\beq T_{g,h}(\tau):=\Tr_{V^\natural_g}(q^{L_0-1}h)\ ,
\eeq where $V^\natural_g$ is the $g$-twisted sector of $V^\natural$. The generalized Monstrous moonshine conjecture, proved in \cite{CarnahanI:2010,CarnahanII:2012,CarnahanIV:2012gx}, states that all such $T_{g,h}$ either vanihs or are Hauptmoduls for some genus zero group $\Gamma_{g,h}$. Notice that $T_h=T_{1,h}$, so that the standard Monstrous moonshine conjecture is a particular case of the generalized one when we restrict to $g=1$.

\bigskip

It is natural to generalize this idea and define the \emph{defect McKay-Thompson series} $T_\CL$ as the torus partition function with the insertion of any (possibly non-invertible) topological defect $\CL$ along a space-like torus. The series is given by
\beq T_\CL(\tau):=\Tr_{V^\natural_g}(q^{L_0-1}\hat\CL)\ ,
\eeq where $\hat\CL:V^\natural\to V^\natural$  is the linear operator associated with $\CL$. Such functions were considered in \cite{Lin:2019hks}, where the series $T_\CL$ for a non-invertible topological defect $\CL$ was also computed. More examples were computed in \cite{Fosbinder-Elkins:2024hff}.

In this section, we provide a general formula for a class of topological defects known as duality defects, and that includes the case studied in \cite{Lin:2019hks} and some of the examples in \cite{Fosbinder-Elkins:2024hff}.  Let $g\in \mathbb{M}$ be a Fricke non-anomalous element of order $N$. This means that the corresponding  McKay-Thompson series $T_g(\tau)$ is invariant, with trivial multiplier, under the group $\Gamma_0(N)+N$, i.e. the extension of the group 
\be \Gamma_0(N):=\left\{\begin{pmatrix}
	a & b\\ c & d
\end{pmatrix}\in SL_2(\ZZ)\mid c\equiv 0\mod N\right\}\ ,
\ee by the Fricke involution $\left(\begin{smallmatrix} 0 & -1/\sqrt{N}\\ 1/\sqrt{N} & 0\end{smallmatrix}
\right)\in SL_2(\RR)$. The fact that the multiplier is trivial implies that the group $\langle g\rangle\cong \ZZ_N$ is not anomalous, so that the orbifold $V^\natural/\langle g\rangle$ is a well-defined holomorphic VOA of central charge $c=24$. Furthermore, one can prove that, for a non-anomalous $g$, the orbifold $V^\natural/\langle g\rangle$ is isomorphic to $V^\natural$ if and only if $T_g$ is invariant under the Fricke involution, i.e.
\beq\label{Frickeinv} T_g(-1/N\tau)=T_g(\tau)\ ,
\eeq see \cite{Paquette:2017xui}.
On general grounds \cite{Frohlich:2006ch}, one expects this `self-orbifold' property to be related to the existence of a topological defect $\CN_g=\CN_g^*$ (a `duality defect' for $g$) such that 
\be \CN_g^2=\sum_{k=0}^{N-1} \CL_{g^k}\ ,\qquad \CN_g\CL_{g^k}=\CN_g=\CL_{g^k}\CN_g
\ee where $\CL_{g^k}$ are the invertible defects related to the automorphisms $g^k$. 
The fusion category generated by the defects $\CN_g$ and $\CL_{g^k}$ is a particular example of a Tambara-Yamagami category, whose properties were studied in \cite{Tambara:1998}. In general, a Tambara-Yamagami category $TY(A)$ is generated by a non-anomalous abelian group $A$ of invertible defects $\CL_a$, $a\in A$, as well as a  duality defect $\CN=\CN^\ast$ such that $\CN^2=\sum_{a\in A}\CL_a$.   Our goal is to provide an explicit general formula for the defect McKay-Thompson series 
\be T_{\CN_g}(\tau)=\Tr_{V^\natural}(q^{L_0-\frac{c}{24}}\hat\CN_g)\ ,
\ee
for all Tambara-Yamagami categories $TY(\ZZ_N)$ associated with a (Fricke, non-anomalous) cyclic subgroup $\ZZ_N$ of the Monster. An analogous classification of the duality defects for the cyclic groups of the $E_8$ lattice VOA was given in \cite{Burbano:2021loy}.

Let $(V^\natural)^{\langle g\rangle}\subset V^\natural$ be the $g$-fixed subVOA of $V^\natural$. For a non-anomalous symmetry  group $\langle g\rangle\cong \ZZ_N$, it is known \cite{Carnahan:2016guf,vanEkeren:2017scl} that $(V^\natural)^{\langle g\rangle}\subset V^\natural$ is strongly rational and has $N^2$ irreducible ordinary modules $V_{n,m}$, $n,m\in \ZZ/N\ZZ$  with conformal weights 
\beq\label{weights} \Delta_{V_{n,m}}=\frac{nm}{N}\mod \ZZ\ ,
\eeq
and satisfying the group-like fusion rules
\beq\label{fusion} V_{i,j}\otimes V_{k,l}\cong  V_{i+k,j+l}\ , \qquad i,j,k,l\in \ZZ/N\ZZ\ .
\eeq
More precisely, $V_{n,m}$ is the $g=e^{2\pi i\frac{m}{N}}$ eigenspace in the $g^n$-twisted sector:
\be V_{n,m}:=\{ v\in V_{g^n}\mid g(v)=e^{2\pi i \frac{m}{N}}v\}\ ,\qquad n,m\in \ZZ/N\ZZ\ ,
\ee where $V_{g^n}$ is the irreducible $g^n$-twisted of $V^\natural$. The action of $g$ on the $g^1$-twisted sector $V_g$ is defined by
\be g_{\rvert V_{g}}=e^{2\pi i L_0}\ ,
\ee while on a generic $g^n$-twisted sector it is defined in such a way that \eqref{fusion} hold. The character of the module $V_{n,m}$ can be written as
\be \Tr_{V_{n,m}}(q^{L_0-1})=\frac{1}{N}\sum_{k=0}^{N-1} e^{-\frac{2\pi i mk}{N}}T_{g^n,g^k}(\tau)\ ,
\ee
in terms of the generalized McKay-Thompson  series
\be T_{g^n,g^k}(\tau)=\Tr_{V_{g^n}}(g^kq^{L_0-1})\ .
\ee The latter can be obtained from the ordinary McKay-Thompson series $T_{e,g^k}\equiv T_{g^k}$ using the modular transformations 
\beq\label{modulartr} T_{g}\left(\frac{a\tau+b}{c\tau+d}\right)=T_{g^c,g^d}(\tau)\ ,
\qquad \begin{pmatrix} a & b\\ c & d\end{pmatrix}\in SL_2(\ZZ)\ ,
\eeq (this formula is modified when the group $\langle g\rangle$ is anomalous). For Fricke elements, by \eqref{Frickeinv} and \eqref{modulartr} we have the relation 
\beq\label{Frickeone} T_{g,e}(\tau)=T_{e,g}(\tau/N)=q^{-1/N}+O(q^{1/N})\ ,\eeq
which implies that the $g$-twisted sector $V_g$ contains a single operator of conformal weight $1-1/N$.

We can decompose both $V^\natural$ and of the orbifold VOA $V^\natural/\langle g\rangle$ in terms of modules of the common subVOA $(V^\natural)^{\langle g\rangle}$. In particular, $V^\natural$ contains all untwisted modules for any eigenvalue of $g$
\be V^\natural=\oplus_{m\in \ZZ/N\ZZ} V_{0,m}\ ,
\ee
while $ V^\natural/\langle g\rangle$ is given by the $g$-invariant twisted and untwisted sectors
\be V^\natural/\langle g\rangle=\oplus_{n\in \ZZ/N\ZZ} V_{n,0}\ .
\ee Notice that, by \eqref{weights}, the conformal weights of $V^\natural/\langle g\rangle$ are all integral, as expected for a non-anomalous $g$.

In \cite{Paquette:2017xui} it was proved that, if $T_g$ is Fricke invariant, then there is an isomorphism of $(V^\natural)^{\langle g\rangle}$ modules
\be f_2:V_{n,m}\stackrel{\cong}{\longrightarrow} V_{m,n}\ ,\qquad n,m\in \ZZ/N\ZZ\ ,
\ee that induces an isomorphism of the VOAs
\be f_2:V^\natural\stackrel{\cong}{\longrightarrow} V^\natural/\langle g\rangle\ .
\ee The linear operator $\hat \CN_g:V^\natural\to V^\natural$ vanishes on the modules $(\hat \CN_g)_{\rvert V_{0,m}}=0$ for $m\neq 0\mod N$, while $(\hat \CN_g)_{\rvert V_{0,0}}=\sqrt{N}f_2$. Therefore, the defect McKay-Thompson series $T_{\CN_g}$ is given by
\beq\label{dualitydefformula} T_{\CN_g}(\tau)=\Tr_{V^\natural}(q^{L_0-\frac{c}{24}}\hat\CN_g)=\sqrt{N}\Tr_{V_{0,0}}(q^{L_0-\frac{c}{24}}f_2)\ .
\eeq
In order to compute  $T_{\CN_g}$, we need some more details about the isomorphism $f_2$.  

Let us consider the lattice VOA $W^L$ based on the even  lattice $L=\sqrt{2N}\ZZ$ of rank $1$. Its modules $W^L_l$,  $l\in L^\vee/L\cong \ZZ/2N\ZZ$, have conformal weights
\beq\label{weights2} \Delta_{W^L_l}= \frac{l^2}{4N}\mod \ZZ\ ,\qquad l\in \ZZ/2N\ZZ\ .
\eeq We consider the product VOA \be \tilde V_{0,0,0}:=W^L\otimes (V^\natural)^{\langle g\rangle}\ee and its modules
\be \tilde V_{l,n,m} :=W^L_{l}\otimes V_{n,m}\ .
\ee  By comparing \eqref{weights} and \eqref{weights2}, it is clear that $\tilde V_{2n,n,-n}$, $n\in \ZZ/N\ZZ$, have integral conformal weights. In particular, by \eqref{Frickeone}, the lowest weight vectors in the modules $V_{1,-1}$ and $V_{-1,1}$ have conformal weight $1-\frac{1}{N}$, while the lowest vectors in $W^L_{2}$ and $W^L_{-2}$ have weight $\frac{1}{N}$. This implies that the lowest weight operators in $V_{2,1,-1}$ and $V_{-2,-1,1}$   are currents $J^+$ and $J^-$ of conformal weight $1$. Therefore, one can consider the  VOA 
\be \tilde V:=\oplus_{n\in \ZZ/N\ZZ} \tilde V_{2n,n,-n}\ ,
\ee that is a simple current extension of  $\tilde V_{0,0,0}$. The $\tilde V$-modules are given by
\be  \oplus_{\substack{n \in \ZZ/N\ZZ}} \tilde V_{2n-l,n,l-n}\ ,\qquad l\in \ZZ/2 N\ZZ\ .
\ee


The operators of weight $1$ in $\tilde V$ are given by the current $H\in \tilde V_{0,0,0}\cong W^L\otimes (V^\natural)^{\langle g\rangle} $ as well as $J^+\in V_{2,1,-1}$ and $J^-\in V_{-2,-1,1}$. Together, they generate a $\hat{su}(2)_N$ affine Kac-Moody algebra. The module $\oplus_{\substack{n \in \ZZ/N\ZZ}} \tilde V_{2n-l,n,l-n}$ is therefore a module also for this $\hat{su}(2)_N$ algebra. In particular, the eigenvalues of the current zero mode $H_0$ on $\tilde V_{2n-l,n,l-n}$ are contained in $2n-l+2N\ZZ$. The $SU(2)$ group generated by the currents' zero modes acts by vertex algebra automorphisms on the extended VOA $\oplus_{n\in \ZZ/N\ZZ} \tilde V_{2n,n,-n}$. This group contains an element $f$ whose adjoint action on the currents is
\be (\mathrm{ad}\ f)(H)= -H\qquad (\mathrm{ad}\ f)(J^\pm)= J^\mp \ .\ee
The action of $f$ on each $\hat{su}(2)_N$-module $\oplus_{\substack{n \in \ZZ/N\ZZ}} \tilde V_{2n-l,n,l-n}$ reverses the $H_0$ eigenvalues, so that it must map each component $\tilde V_{2n-l,n,l-n}$ to $\tilde V_{l-2n,l-n, n}$. In particular, $f$ restricts to an automorphism of the subalgebra $\tilde V_{0,0,0}=W^L\otimes  (V^\natural)^{\langle g\rangle}$, and it is easy to see that $f_{\rvert \tilde V_{0,0,0}}$ factorizes as a tensor product $f_{\rvert \tilde V_{0,0,0}}=f_1\otimes f_2$ of automorphisms $f_1\in \Aut(W^L)$ and $f_2\in \Aut((V^\natural)^{\langle g\rangle})$. Because $f_1\otimes f_2$ induces an isomorphism 
\be f_1\otimes f_2:\tilde V_{2n-l,n,l-n} \stackrel{\cong}{\longrightarrow} \tilde V_{l-2n,l-n, n}\ , \qquad l\in \ZZ/2N\ZZ,\ n\in \ZZ/N\ZZ\ ,
\ee of $W^L\otimes (V^\natural)^{\langle g\rangle}$ modules, it follows that $f_2$ induces an isomorphism 
\be f_2:V_{n,m} \stackrel{\cong}{\longrightarrow} V_{m,n}\ ,\qquad n,m\in \ZZ/N\ZZ\ ,
\ee of $(V^\natural)^{\langle g\rangle}$-modules, compatible with fusion. By applying the map $f_2$ to $V^\natural =\oplus_{m\in \ZZ/N\ZZ} V_{0,m}$ we obtain an isomorphism with $V^\natural/\langle g\rangle =\oplus_{n\in \ZZ/N\ZZ} V_{n,0}$. Therefore, $f_2$ is the automorphism of $(V^\natural)^g$ appearing in \eqref{dualitydefformula}.

In order to compute the trace $\Tr_{V_{0,0}}(q^{L_0-1}f_2)$, it is useful to first consider
\beq\label{traceVtilde} \Tr_{\tilde V}(fq^{L_0-\frac{25}{24}}).
\eeq
Let us decompose $\tilde V$ into eigenspaces for the current zero mode $H_0$:
\be \tilde V=\bigoplus_{n\in \ZZ} U_{\frac{n}{\sqrt{2N}}}\otimes V_{n,-n}\ ,
\ee where $U_\lambda$, $\lambda\in \RR$ denotes a module for the Heisenberg algebra generated by the modes of the $u(1)$ current $H(z)$. Now, because of the action $(\mathrm{ad}\ f)(H)= -H$, the isomorphism $f$ must map each $U_{\frac{n}{\sqrt{2N}}}\otimes V_{n,-n}$ to $U_{\frac{-n}{\sqrt{2N}}}\otimes V_{-n,n}$, so that the only sector that contributes to the trace \eqref{traceVtilde} is the one with $n=0$.  Thus, we have
\be \Tr_{\tilde V}(fq^{L_0-\frac{25}{24}})=\Tr_{U_{0}\otimes V_{0,0}}(fq^{L_0-\frac{25}{24}})=\Tr_{U_{0}}(f_1q^{L_0-\frac{1}{24}})\Tr_{ V_{0,0}}(f_2q^{L_0-\frac{24}{24}})\ .
\ee Notice, in particular, that $f_1\in \Aut(W^L)$ preserves the Heisenberg subalgebra $U_{0}\subset W_L$. We obtain
\be \Tr_{ V_{0,0}}(f_2q^{L_0-1})=\frac{\Tr_{\tilde V}(fq^{L_0-\frac{25}{24}})}{\Tr_{U_{0}}(f_1q^{L_0-\frac{1}{24}})}\ .
\ee Now, we compute the numerator and denominator on the right-hand side. The denominator is simply given by
\be \Tr_{U_{0}}(f_1q^{L_0-\frac{1}{24}})=q^{-\frac{1}{24}}\frac{1}{\prod_{n=1}^\infty (1+q^n)}=\frac{\eta(\tau)}{\eta(2\tau)}\ .
\ee To compute the numerator, we use the fact that $f$ is conjugate, within the group $SU(2)$, of an element $f'=e^{\pi i H_0}$ with adjoint action
\be (\mathrm{ad}\ f')(H)= H\qquad (\mathrm{ad}\ f')(J^\pm)=- J^\pm \ ,\ee so that
\be \Tr_{\tilde V}(fq^{L_0-\frac{25}{24}})=\Tr_{\tilde V}(f'q^{L_0-\frac{25}{24}}).
\ee We have
\be \Tr_{\tilde V}(f'q^{L_0-\frac{25}{24}})=\sum_{n\in \ZZ/N\ZZ}  \Tr_{\tilde V_{2n,n,-n}}(f'q^{L_0-\frac{25}{24}})=\sum_{n\in \ZZ/N\ZZ} \Tr_{W^L_{2n}}(f'q^{L_0-\frac{1}{24}})\Tr_{V_{n,-n}}(q^{L_0-\frac{24}{24}})\ .
\ee
It is easy to compute
\be \Tr_{W^L_{2n}}(f'q^{L_0-\frac{1}{24}})=\frac{\Theta_{\frac{2n}{\sqrt{2N}}+L}(\tau,\frac{1}{2}\sqrt{\frac{N}{2}})}{\eta(\tau)}\ ,
\ee where 
\be \Theta_{\frac{l}{\sqrt{2N}}+L}(\tau,z):=\sum_{k\in \ZZ} e^{2\pi iz(\frac{l}{\sqrt{2N}}+k\sqrt{2N})}q^{\frac{(l+2kN)^2}{4N}}\ .
\ee
is the theta series of the translated lattice $\frac{l}{\sqrt{2N}}+L=\frac{l}{\sqrt{2N}}+\sqrt{2N}\ZZ$.
In particular,
\begin{align} \Theta_{\frac{2n}{\sqrt{2N}}+L}(\tau,\tfrac{1}{2}\sqrt{\tfrac{N}{2}})&=\sum_{k\in \ZZ} (-1)^{n+kN}q^{\frac{(n+kN)^2}{N}}=q^{\frac{n^2}{N}}(-1)^n\sum_{k\in \ZZ} (-1)^{kN}q^{k^2N+2nk}\notag\\
&=q^{\frac{n^2}{N}}(-1)^n\theta_3(2N\tau,2n\tau+N/2)\ ,
\end{align} in terms of the Jacobi theta function $\theta_3(\tau,z)=\sum_{k\in \ZZ}q^{\frac{k^2}{2}}e^{2\pi i z k}$.
We conclude that
\beq T_{\CN_g}(\tau)=\sqrt{N}\frac{\eta(2\tau)}{\eta(\tau)^2}\sum_{n\in \ZZ/N\ZZ} \Theta_{\frac{2n}{\sqrt{2N}}+L}(\tau,\tfrac{1}{2}\sqrt{\tfrac{N}{2}})\Tr_{V_{n,-n}}(q^{L_0-1})\ .
\eeq
For example, for an element $g$ in class $2A$ of the Monster, we have
\be \Tr_{V_{0,0}}(q^{L_0-1})=\frac{T_{e,e}(\tau)+T_{e,g}(\tau)}{2}=\frac{J(\tau)+T_{2A}(\tau)}{2}
\ee
\be \Tr_{V_{1,1}}(q^{L_0-1})=\frac{T_{g,e}(\tau)-T_{g,g}(\tau)}{2}=\frac{T_{2A}(\frac{\tau}{2})-T_{2A}(\frac{\tau+1}{2})}{2}
\ee
so that
\begin{align} T_{\CN_{2A}}(\tau)&=\sqrt{2}\frac{\eta(2\tau)}{\eta(\tau)^2}\Bigl(\theta_3(4\tau)\frac{J(\tau)
	+T_{2A}(\tau)}{2}-\theta_2(4\tau)\frac{T_{2A}(\frac{\tau}{2})-T_{2A}(\frac{\tau+1}{2})}{2}\Bigr)\notag\\
&=\sqrt{2}\Bigl(\frac{1}{q}+ 91886 q + 8498776 q^2 + 301112552 q^3 + 6338608848 q^4 + \ldots\Bigr)\ .
\end{align} The result coincides with the one given in \cite{Lin:2019hks}.
Similarly, for $g$ in class $3A$, we have
\be T_{\CN_{3A}}(\tau)=\sqrt{3}\Bigl(\frac{1}{q} + 48808 q + 3802016 q^2 + 118187964 q^3 + 2227681632 q^4  +\ldots \Bigr)\ .
\ee

\section{Moonshine, BKM algebras, and string theory}\label{s:Moonshine}

The discussions in sections \ref{s:TDLreview} and \ref{s:Monstrous} and the results in \cite{Lin:2019hks} and \cite{Fosbinder-Elkins:2024hff}, yield the following question:
\begin{itemize}
	\item[Problem 1:] {\it Determine the fusion category $\calC_{Vir}(V^\natural)$ of topological defects of $V^\natural$ commuting with the Virasoro algebra and determine the defect Mc-Kay-Thompson series $T_\CL$ for all objects in this category.}
\end{itemize}
Unfortunately, the resolution of this problem seems to be out of reach at the moment. The main reason is that the Virasoro algebra at $c=24$ is not rational. This means that, strictly speaking, the definition we proposed in section \ref{s:TDLreview} cannot be used in this case. In particular, because of the lack of rationality, one expects the putative `fusion category' of defects preserving only $Vir$ to include infinitely many simple objects. 

One way to approach this problem would be to study the fusion categories $\calC_W(V^\natural)$ for all strongly rational subVOAs $W$ that are conformally embedded in $V^\natural$. While finding all such embedded subalgebras $W$ is by itself a very difficult open problem, it would provide at least a well posed mathematical question.

Even if we were able to perform such a classification, we have reasons to expect that many simple defects in $\calC_{Vir}(V^\natural)$ would not be included in any $\calC_W(V^\natural)$ for any rational $W$.  For example, the fact that $V^\natural$ can be obtained as a $\ZZ_2$-orbifold $V^\natural=V_\Lambda/\langle \iota\rangle$ of the lattice VOA $V_\Lambda$ based on the Leech lattice $\Lambda$, suggests that there should be a continuum of simple topological defects $\CL_{\vec\theta}$ parametrized by a collection of real parameter $\vec\theta=(\theta_1,\theta_2,\ldots)$. In order to see this, recall that the group of automorphisms $\Aut(V_\Lambda)$ contains a subgroup of inner automorphisms $U(1)^{24}$, generated by the current zero modes. Let $\CW_{\vec\theta}$, $\vec\theta\in (\RR/2\pi\ZZ)^{24}$ denote the invertible topological defect of $V_\Lambda$ corresponding to such inner automorphisms. Taking the orbifold of $V_\Lambda$ by the involution $\iota$ projects all currents of $V_\Lambda$ out, and only the finite subgroup $\ZZ_2^{24}\subset U(1)^{24}$ of symmetries commuting with $\iota$ survives as a collection of invertible topological defects of $V^\natural$. However, for each $\vec\theta\in (\RR/2\pi\ZZ)^{24}$, the superposition
\be \CW_{\vec\theta}+\CW_{-\vec\theta}
\ee commutes with $\iota$, and gives rise to a non-invertible topological defect $\CL_{\vec\theta}$ in $V^\natural$. As far as we know, infinite fusion categories containing such continuous families of simple objects have not been studied in detail so far. For generic values of the continuous parameters $\vec\theta$, we expect the fusion products $(\CL_{\vec\theta})^n$ of the defect $\CL_{\vec\theta}$ with itself to give rise to new simple defects at every $n$ -- this is what happens for the invertible defect $\CW_{\vec{\theta}}$ in $V_\Lambda$ at generic $\vec\theta$. This implies that $\CL_{\vec\theta}$ cannot be contained in any fusion category with finitely many simple objects, and in particular in any  $\calC_W(V^\natural)$ for rational $W$.

The appearance of continuous families of defects in a VOA is not surprising, given that this can happen already at the level of invertible defects.  In general, it is known that the group of automorphisms $\Aut(V)$ of a holomorphic VOA $V$ contains a normal Lie group $\Aut_{in}(V)\subseteq \Aut(V)$ of inner automorphisms generated by the current zero modes, so that the quotient $\Aut(V)/\Aut_{in}(V)$ (the group of outer automorphisms) is finite. It is natural to wonder whether some analogous result holds for fusion categories of topological defects. In fact, physics arguments based on a generalization of Noether theorem \cite{Thorngren:2021yso}, suggest that the existence of a continuous $1$-parameter family of topological defects $\CL_\theta$, satisfying $\CL_\theta\CL_{\theta'}=\CL_{\theta+\theta'}$ and $\CL_\theta^\ast=\CL_{-\theta}$, is necessarily associated with a conserved current $J(z)\in \Hh_{\CL'}$, where $\CL'$ is a subobject of $\CL_\theta\CL_\theta^*$ for all $\theta$. Furthermore, any two defects $\CL_\theta$ and $\CL_{\theta'}$ in the family can be obtained from each other by inserting a suitable `exponentiation' $e^{\int_\gamma J(z)dz}$ on the support $\gamma$ of the defect. These arguments show that continuous families of topological defects are, in a sense, the direct analog of a Lie group of inner automorphisms. If a proper notion of `quotienting out the continuous families of defect' existed, analogous to the quotient $\Aut(V)/\Aut_{in}(V)$ by the group of inner automorphisms, then one could focus in determining the `quotient category', which might be a feasible task. Unfortunately, we are not sure whether one can make sense of these speculations.

\bigskip

Finally, even if we were able to solve problem 1 in a satisfactory way, it is unlikely that the answer would provide a generalization of the Monstrous moonshine conjecture. Indeed, the results of \cite{Lin:2019hks} and \cite{Fosbinder-Elkins:2024hff} show that some McKay-Thompson series $T_\CL$ associated with non-invertible defects are not invariant under any genus zero subgroup of $SL_2(\RR)$. Even when the invariance subgroup is genus zero, the series $T_\CL$ might not necessarily be the Hauptmodul. These remarks suggest a second open question:

\begin{itemize}
	\item [Problem 2:] {\it Determine the largest fusion category (or categories) $\calC^{MS}_{Vir}(V^\natural)$ of topological defects of $V^\natural$ exhibiting `Moonshine properties', i.e. such that for all simple $\CL\in\calC^{MS}_{Vir}(V^\natural)$ the McKay-Thompson series 
		\beq\label{McKay} T_\CL(\tau)=\sum_{n=0}^\infty \Tr_{V^\natural_n}(\hat\CL)q^{n-1}=\langle \CL\rangle q^{-1}+O(q)\eeq is a Hauptmodul for a discrete genus zero subgroup  $\Gamma_\CL\subset SL_2(\RR)$  (and explain why the category has this property!).}
\end{itemize} A priori, we do not know whether a single fusion category containing all topological defects of `Moonshine type' exist, so we have to allow for the possibility of having different maximal fusion categories with this property. We are particularly interested in categories that contain all invertible defects $\CL_g\in \mathbb{M}$. Notice that we cannot require the McKay-Thompson series $T_\CL$ to be a Hauptmodul when the defect $\CL$ is non-simple: for example, this is not true even for sums $\CL=\CL_g+\CL_h$ of invertible topological defects with $g,h\in \mathbb{M}$.

We have been quite vague with the definition of the invariance group $\Gamma_\CL$. By definition, it is always true that $T_\CL(\tau)$ is invariant under  $\tau\to\tau+k$ if and only if $k\in \ZZ$.  One might want to put some further constraints on $\Gamma_\CL$, for example requiring it to be  commensurable with $SL_2(\ZZ)$ (i.e. such that $\Gamma_\CL\cap SL_2(\ZZ)$ has finite index both in $\Gamma_\CL$ and in $SL_2(\ZZ)$) or a congruence subgroup, i.e. to contain the principal congruence group 
$$\Gamma(N)=\{\begin{pmatrix}
a & b\\ c & d
\end{pmatrix}\in SL_2(\ZZ)\mid a,d\equiv 1\mod N,\ b,c\equiv 0\mod N\}$$
at some finite level $N$. In the latter case, $\Gamma_{\CL}$ would be a group of `moonshine type', as defined in \cite{ConwayNorton1979}. Because there is a finite number of genus zero congruence subgroups of $SL_2(\RR)$, then there is a finite number of possible defect McKay-Thompson series. This suggests that the category of defects in Problem 2 might contain a finite number of simple objects, and therefore much more manageable than the one in Problem 1. Notice, however, that in principle there might be many (maybe even infinitely many?) distinct simple defects with the same McKay-Thompson series.

A sufficient condition for the group $\Gamma_\CL$ to be a congruence subgroup is that the subalgebra $(V^\natural)^{\CL}$ preserved by $\CL$ is rational. In this case, we can consider $\CL$ as an object in the fusion category of topological defects preserving $(V^\natural)^\CL$; such category contains only finitely many simple defects. Each McKay-Thompson series in this category is a linear combination of characters for the algebra $(V^\natural)^\CL$, so it is necessarily invariant under some principal congruence subgroup $\Gamma(N)$ for some level $N$.

By definition \eqref{McKay}, the series $T_\CL$ are not normalized when $\CL$ is not invertible. Furthermore, their Fourier coefficients are not necessarily rational, as the examples in section \ref{s:Monstrous} show. For a duality defect $\CN_g$, and more generally for defects belonging to a Tambara-Yamagami fusion category, the rescaling 
\be \frac{1}{\langle \CN_g\rangle}T_{\CN_g}(\tau)=q^{-1}+\ldots\ ,
\ee gives a modular function with rational (in fact, integral) coefficients and normalized so that the pole at $\infty$ has residue $1$. For more exotic fusion categories, the corresponding McKay-Thompson series is expected to be genuinely irrational, i.e. not just an irrational rescaling of a rational function. Thus, considering defect McKay-Thompson series might provide examples of irrational Hauptmoduls.

\bigskip

The possibility of modular functions $T_\CL$ with irrational coefficients is in sharp contrast with the results with the series $T_g$ associated with $g\in\Aut(V^\natural)$, that are known to have integral coefficients for all $g$. This property can be seen as a consequence of the existence of  a self-dual integral form for the Monster module $V^\natural$ \cite{Carnahan2019}. This means that there exists a VOA $V^\natural_\ZZ$ defined over the ring of integers with a self-dual invariant non-degenerate bilinear form, and such that $V^\natural_\ZZ\otimes \CC$ reproduces the usual FLM VOA $V^\natural$ over the complex numbers. Furthermore, the Monster group is a symmetry of this integral form preserving the bilinear form. Based on this observation, Carnahan and Urano formulate a conjectured encompassing both Conway and Norton's Monstrous moonshine and Borcherds-Ryba modular moonshine \cite{CarnahanUrano2024}. 

This leads us to consider the following question:
\begin{itemize}
	\item[Problem 3:] Determine the largest fusion category $\calC_{Vir}(V^\natural_\ZZ)$ of topological defects of $V^\natural$ that restrict to $V^\natural_\ZZ$, i.e. such that the linear map $\hat\CL:V^\natural \to V^\natural$ is the extension by $\CC$-linearity of a $\ZZ$-linear endomorphism  $\hat\CL_\ZZ:V^\natural_\ZZ\to V^\natural_\ZZ$.
\end{itemize} Because a topological defect acts on the vacuum $|0\rangle$ of $V^\natural$  by a rescaling $\hat\CL|0\rangle =\langle\CL\rangle|0\rangle$, it follows that all defects in $\calC_{Vir}(V^\natural_\ZZ)$ must have integral quantum dimension -- it is an integral fusion category.

Notice that Carnahan and Urano's proposal would extend very naturally to the fusion category $\calC_{Vir}(V^\natural_\ZZ)$, while no such generalization seems possible for more general topological defects. Based on this observation, it is very tempting to speculate that the category $\calC_{Vir}(V^\natural_\ZZ)$ is closely related (or, possibly, coincides?) with one of the Moonshine categories $\calC^{MS}_{Vir}(V^\natural)$ in Problem 2. 

As we will see in section \ref{s:Conway}, in the superMoonshine case of the vertex operator superalgebra $V^{f\natural}$, there is a suitable even self-dual lattice $\Lambda$ in the space of $24$ Ramond ground states, such that for every topological defect $\CL$  commuting with the $\CN=1$ superVirasoro algebra, the linear map $\hat\CL$ induces a lattice endomorphism. We do not know whether there is an integral form $V_\ZZ^{f\natural}$ of $V^{f\natural}$ with an invariant self-dual integral bilinear form;  if it exists, it is natural to conjecture that the lattice $\Lambda$ is contained in the twisted module of $V_\ZZ^{f\natural}$. If a suitable $V^{f\natural}_\ZZ$ the analogous version of Problem 3 could be formulated for the superMoonshine module.

\bigskip

Problems 2 and 3 have the potential to shed a new light on the meaning of the Moonshine conjectures, especially the genus zero properties of McKay-Thompson series. In particular, if (some version of) the  proof of Monstrous moonshine can be extended to a fusion category of topological defects, then we would expect such category to be of `Moonshine type', as described in problem 2.

Borcherds' proof of Monstrous moonshine \cite{Borcherds1992} is based on the construction of an infinite dimensional Lie algebra $\mathfrak{m}$ (a Borcherds-Kac-Moody algebra) defined in terms of the FLM module $V^{\natural}$, and that inherits the action of the Monster group as a group of algebra automorphisms. More into detail, the generators of the BKM algebras correspond to the physical states in a chiral bosonic string theory based on the holomorphic CFT $V^\natural\otimes V_{II_{1,1}}$, where  $V_{II_{1,1}}$ is the lattice vertex algebra based on the indefinite even unimodular lattice of signature $(1,1)$. Analogous constructions, involving different BKM algebras $\mathfrak{m}_g$, with $g\in \mathbb{M}$,  has been used for the proof of generalized moonshine \cite{CarnahanIV:2012gx}. The identification of the McKay-Thompson series with the Hauptmoduls of some genus zero subgroups of $SL_2(\RR)$ is proved using some `replica identities' that are the consequences of (twisted) denominator identities for the BKM algebras.

While, by construction, every automorphism $g$ of the Monster module $V^\natural$ induces an automorphism of the BKM algebra $\mathfrak{m}$, it is not clear what the analogous statement would be for topological defects.

\begin{itemize}
	\item[Problem 4.] Suppose that $V^\natural$ admits a topological defect $\mathcal{L}$. Describe the implications of the existence of $\mathcal{L}$ for the BKM algebra $\mathfrak{m}$. 
\end{itemize}

In \cite{Paquette:2016xoo,Paquette:2017xui}, a slightly different construction of the Monstrous BKM algebra was proposed, based on a non-chiral heterotic string compactified on $(V^\natural\otimes \overline{V^{f\natural}})\times S^1$. Here, $\overline{V^{f\natural}}$ is the anti-holomorphic version of the Conway module SVOA $V^{f\natural}$ (see section \ref{s:Conway}), and $S^1$ is the superconformal field theory of central charges $(c,\tilde c)=(1,3/2)$ corresponding to a free boson on a circle $S^1$ and a antiholomorphic free fermion. Analogous string theory models can be defined on suitable `CHL orbifolds' of the $(V^\natural\otimes \overline{V^{f\natural}})\times S^1$ CFT by symmetries $(g,\delta)$, where $g\in \mathbb{M}$ is an automorphism of $V^\natural$ of order $N$ and $\delta$ is a shift along the circle $S^1$ by $1/N$ of a period. This provides a uniform physics construction of the BKM algebras that are related to generalized Monstrous Moonshine. In these string models, the McKay-Thompson series $T_g$ are related to certain supersymmetric indices counting (second quantized) BPS states, and the invariance subgroups $\Gamma_g$ are interpreted as T-duality groups. 
Once topological defects are considered, the following question arises naturally:
\begin{itemize}
	\item[Problem 5.] Generalize the string theory construction of \cite{Paquette:2016xoo,Paquette:2017xui} so as to include topological defects of $V^\natural$.
\end{itemize}
This problem is closely related to an open problem in string theory. Indeed, the construction of \cite{Paquette:2016xoo, Paquette:2017xui} is based on the presence of string dualities, whose existence is due to symmetries of the two-dimensional CFT that defines the string model.  In general, the dynamics of a (super)string moving in a space-time with some given topology and geometry (the target space) is described, at least perturbatively, by a two dimensional conformal field theory on the world-sheet of the string. The physical states of the string in space-time can be obtained as the cohomology classes for a suitable BRST operator on the worldsheet CFT. When the two-dimensional CFT has a group of (ordinary) global symmetries preserving the BRST operator, this leads to selection rules for the string amplitudes that hold at all orders in perturbation theory. Usually, such selection rules are associated with a gauge symmetry for the physical theory in the target space. This statement has no obvious generalization to the case of a non-invertible topological defect $\mathcal{L}$ in the two-dimensional CFT on the worldsheet. Even when $\mathcal{L}$ preserves the BRST operator, the selection rules for the string amplitudes seem to hold only at tree level (i.e. on the sphere), while they are in general broken at higher loop level (i.e. for worldsheets with genus higher than $0$). See for example \cite{Heckman:2024obe,Kaidi:2024wio} for discussions. It is not clear whether the existence of $\mathcal{L}$ has any (less obvious) consequences for the space-time theory.

\section{Topological defects in Conway module}\label{s:Conway}

While finding all topological defects of $V^\natural$ preserving the Virasoro algebra seems impossible at the moment, some partial result can be obtained for an analogous problem in case of holomorphic super-vertex operator algebras (SVOA).

Let $V$ be a (rational, $C_2$-cofinite, of CFT-type) SVOA with central charge $c=12$. The SVOA $V$ is called holomorphic (or self-dual) if the only irreducible modules is $V$ itself, up to isomorphism. It was proved in \cite{Creutzig:2017fuk} that there are only three self-dual SVOAs at $c=12$, namely:
\begin{itemize}
	\item The SVOA $F_{24}$ generated by $24$ free fermions. This is also the lattice SVOA based on the odd unimodular lattice $\ZZ^{12}$.
	\item $V^{fE_8}$, the product of the $E_8$ lattice VOA and $8$ free fermions; equivalently, it is the lattice SVOA based on $E_8\oplus\ZZ_4$;
	\item the Conway module $V^{f\natural}$ studied in \cite{Duncan:2006}, which is the only self-dual SVOA with no operators of weight $1/2$. It is the SVOA based on the lattice
	\be D_{12}^+:=\{\frac{1}{2}(x_1,\ldots,x_{12})\in (\frac{1}{2}\ZZ)^{12}\mid x_i\equiv x_j\mod 2,\ \sum_i x_i\in 4\ZZ \}\ .
	\ee
\end{itemize}
Let $V$ be one of these three SVOAs. Then, $V$ admits a unique (up to isomorphism) canonically twisted module  $V_{tw}$, i.e. a module twisted by the fermion number $(-1)^F$. $V_{tw}$ is itself a superspace, in the sense that it admits a $\ZZ_2$-grading by $(-1)^F$ that is compatible with the one on $V$.\footnote{For $V^{f\natural}$, there are actually two inequivalent choices for the action of $(-1)^F$ on $V^{f\natural}_{tw}$, one where the $24$ ground states are even and one where they are odd. We will choose the former.}

Each of these SVOA admit a (not necessarily unique) supercurrent $\tau(z)$ extending the Virasoro algebra $Vir_{c=12}$ to $\CN=1$ superVirasoro $SVir_{c=12}$. In particular, $V$ is a module for the Neveu-Schwartz (NS) $\CN=1$ superconformal algebra and $V_{tw}$ a module for the Ramond (R) algebra. For these reasons, following the physicists conventions, we often refer to $V$ and $V_{tw}$ as the NS and R sector, respectively.


\bigskip

With each theory $V$, one can associate four torus partition functions $Z^+_{NS}$, $Z^-_{NS}$, $Z^+_{R}$, $Z^-_{R}$, corresponding to the four choices of spin structure. They are given by
\begin{align*}
Z^\pm_{NS}(\tau)&:=\Tr_{V}(q^{L_0-\frac{c}{24}}(\pm 1)^F)\\
Z^\pm_{R}(\tau)&:=\Tr_{V_{tw}}(q^{L_0-\frac{c}{24}}(\pm 1)^F)\ .
\end{align*} It is useful to organize them into a $4$-component vector
\be \zZ:=(Z^+_{NS}, Z^-_{NS}, Z^+_{R}, Z^-_{R})^t
\ee that transforms as a vector-valued modular function under $SL_2(\ZZ)$ transformations
\be \zZ(\tau+1)=\rho(T)\zZ(\tau)\ ,\qquad\qquad \zZ(-1/\tau)=\rho(S)\zZ(\tau)
\ee where the $SL_2(\ZZ)$-generators $T=\left(\begin{smallmatrix}
1 & 1\\ 0 &1
\end{smallmatrix}\right)$ and $S=\left(\begin{smallmatrix}
0 & -1 \\1 &0
\end{smallmatrix}\right)$ are represented by the matrices\footnote{For a generic central charge $c\in \frac{1}{2}\ZZ$, the minus signs in $\rho(T)$ should be replaced by $e^{-2\pi i c/24}$.}
\beq\label{modularspin} \rho(T)=\begin{pmatrix}
	0 & -1 & 0 & 0\\
	-1 & 0 & 0 & 0\\
	0 & 0 & 1 & 0\\
	0 & 0 & 0 & 1
\end{pmatrix}\ ,\qquad \rho(S)=\begin{pmatrix}
	1 & 0 & 0 & 0\\
	0 & 0 & 1 & 0\\
	0 & 1 & 0 & 0\\
	0 & 0 & 0 & 1
\end{pmatrix}
\eeq
Notice that $Z^-_{R}(\tau)$ is modular invariant, and in fact it is a constant that equals the Witten index
\be Z^-_{R}(\tau)=\Tr_{V_{tw}(1/2)}((-1)^F)\ ,
\ee counting the number of bosons minus fermions in the space $V_{tw}(1/2)\subset V_{tw}$ of Ramond ground states. This follows because for unitary representations of the Ramond $\CN=1$ algebra, each $L_0$-eigenspace contains the same number of bosons and fermions, except possibly for $L_0=c/24=1/2$, which is the lowest possible $L_0$-eigenvalue.  The Witten index $Z^-_{R}(\tau)$ equals $24$ for $V^{f\natural}$, while it vanishes for the other two self-dual SVOAs at $c=12$.

\bigskip

Let us consider the category of topological defects for the SVOA $V$ that commute with a given $\CN=1$ superconformal algebra $SVir_{c=12}$ and are well-behaved with respect to the fermion number $(-1)^F$. More specifically, we consider defects $\CL$ satisfying the following conditions:
\begin{enumerate}
	\item The linear map $\hat\CL:V\to V$ commutes with $(-1)^F$ and with $SVir_{c=12}$. The same holds for the induced linear map $\hat\CL:V_{tw}\to V_{tw}$.
	\item The $\ZZ_2$-grading by $(-1)^F$ can be extended to the $\CL$-twisted sector  $V_{\CL,NS}\equiv V_\CL$ and to the $(-1)^F\CL$-twisted sector $V_{\CL,R}\equiv V_{(-1)^F\CL}$, in a way compatible with the $\ZZ_2$-grading on $V$.\end{enumerate}
If properties 1 and 2 hold, then we have a well-defined  $\CL$-twining partition function $\zZ^\CL:=(Z^{\CL,+}_{NS}, Z^{\CL,-}_{NS}, Z^{\CL,+}_{R}, Z^{\CL,-}_{R})^t$ with components
\begin{align*}
Z^{\CL,\pm}_{NS}(\tau)&:=\Tr_{V}(q^{L_0-\frac{c}{24}}(\pm 1)^F\hat\CL)\\
Z^{\CL,\pm}_{R}(\tau)&:=\Tr_{V_{tw}}(q^{L_0-\frac{c}{24}}(\pm 1)^F\hat\CL)\ ,
\end{align*}
and a $\CL$-twisted partition function $\zZ_\CL:=(Z^+_{\CL,NS}, Z^-_{\CL,NS}, Z^+_{\CL,R}, Z^-_{\CL,R})^t$ with components
\begin{align*}
Z^\pm_{\CL,NS}(\tau)&:=\Tr_{V_\CL}(q^{L_0-\frac{c}{24}}(\pm 1)^F)\\
Z^\pm_{\CL,R}(\tau)&:=\Tr_{V_{(-1)^F\CL}}(q^{L_0-\frac{c}{24}}(\pm 1)^F)\ .
\end{align*} We also require the following:
\begin{enumerate}\setcounter{enumi}{2}
	\item $\zZ^\CL$ and $\zZ_\CL$ transform into one another under modular S-transformations
	\beq\label{Stransftwin} \zZ_\CL(-1/\tau)=\rho(S)\zZ^\CL(\tau)
	\eeq where $\rho(S)$ is the same as in eq.\eqref{modularspin}. 
\end{enumerate}
Property $3$ should be automatically true for defects $\CL$ that preserve a rational subalgebra of $V$; however, a generic $\CL\in \calC_{SVir}(V)$ is expected to preserve only $SVir_{c=12}$, which is not rational.
Properties 1 and 2 imply that  $V_{\CL,NS}$ and  $V_{\CL,R}$ carry (unitary) representations of, respectively, the Neveu-Scwharz and Ramond $\CN=1$ superVirasoro algebra at $c=12$. An argument similar to the one given for the Witten index $Z_R^-$ imply that $Z^{\CL,-}_{R}$ and $Z^-_{\CL,R}$ must be independent of $\tau$
\be Z^{\CL,-}_{R}=\Tr_{V_{tw}(1/2)}((-1)^F\hat\CL)\ ,\qquad Z^{-}_{\CL,R}=\Tr_{V_{(-1)^F\CL}(1/2)}((-1)^F)\ .
\ee Furthermore, by \eqref{Stransftwin}, they must be equal $Z^{\CL,-}_{R}=Z^-_{\CL,R}$. It follows that 
\be \Tr_{V_{tw}(1/2)}((-1)^F\hat\CL)=\Tr_{V_{(-1)^F\CL}(1/2)}((-1)^F)\in \ZZ\ ,\ee
and the latter is clearly an integral number.

Let $\calC_{SVir}(V)$ be a fusion category of topological defects satisfying properties 1, 2, 3 above, and let $G\subseteq \Aut(V)$ be the group of symmetries  generated by all invertible defects in $\calC_{SVir}(V)$. Then, any $\CL\in \calC_{SVir}(V)$ must satisfy
\beq\label{integral} \Tr_{V_{tw}(1/2)}((-1)^Fg\hat\CL)\in \ZZ\ ,\qquad \forall g\in G\ .
\eeq 

\bigskip

Let us now focus on the case $V=V^{f\natural}$. The group of automorphisms $\Aut(V^{f\natural})\cong Spin(24)$ is generated by the zero modes of the currents and acts faithfully on $V^{f\natural}\oplus V_{tw}^{f\natural}$. The centre $\langle (-1)^F,\iota\rangle\cong \ZZ_2\times\ZZ_2$ of $Spin(24)$ is generated  by the fermion number and by an involution $\iota$ acting trivially on $V^{f\natural}$ and by $-1$ on $V^{f\natural}_{tw}$. Under the projection $Spin(24)\to SO(24)$, the involutions $(-1)^F$ and $\iota$ map, respectively, to $1$ and $-1$.

It was proved in \cite{Duncan:2006} that $V^{f\natural}$ admits a unique (up to automorphisms) $\CN=1$ supercurrent $\tau(z)$, and that the subgroup $G\subset Aut(V^{f\natural})$ fixing $\tau(z)$ is isomorphic to the Conway group $G\cong Co_0\subset Spin(24)$. The centre $\ZZ_2\subset Co_0$ is $\langle \iota\rangle$ and acts non-trivially only on the twisted module $V^{f\natural}_{tw}$; thus, strictly speaking, the group acting faithfully on $V^{f\natural}$ is $Co_0/\ZZ_2\cong Co_1$.

All defects $\CL_g$, $g\in G\cong Co_0$, satisfy properties 1,2, and 3 above, so that we can consider a category $\calC_{SVir}(V^{f\natural})$ containing all such invertible defects. For $V=V^{f\natural}$, we can be very explicit about all possible linear maps $\hat\CL:V_{tw}(1/2)\to V_{tw}(1/2)$ satisfying the condition \eqref{integral}. It is known that $G\cong Co_0$ acts on the space of Ramond ground states $V_{tw}(1/2)\cong \RR^{24}$ in its irreducible $24$-dimensional representation. The action of $Co_0$ in this representation preserves a lattice $\Lambda \subset \RR^{24}$ isomorphic to the Leech lattice. 
%
%
%
In \cite{Angius:2024xxx}, we prove the following theorem:
\begin{theorem}\label{th:main}
	Let $V\cong \RR^{24}$ be a $24$-dimensional real vector space with Euclidean metric, $\Lambda\subset V$ a copy of the Leech lattice, $\Aut(\Lambda)\subset SO(V)$ be its group of automorphisms (so that $\Aut(\Lambda)\cong Co_0$), and  $\hat\CL: V\to V$ be a $\RR$-linear map. Then, the following are equivalent:
	\begin{enumerate}
		\item $\Tr_V(\hat\CL g)\in \ZZ\ ,\qquad \forall g\in \Aut(\Lambda)\cong Co_0$.
		\item $\hat\CL$ maps vectors of $\Lambda$ into vectors of $\Lambda$, i.e. $\hat\CL(v)\in \Lambda$ for all $v\in \Lambda$.
		\item $\hat\CL=\sum_{g\in \Aut(\Lambda)} n(g) g$ for some $n(g)\in \ZZ$.
	\end{enumerate}
\end{theorem}

Let us sketch the idea of the proof; see \cite{Angius:2024xxx} for more details. It is obvious that $(2)$ implies $(1)$. Indeed, one can choose a set of generators of the lattice $\Lambda$ as a basis for $V$, and with respect to this basis both $\hat\CL$ and $g$ are represented by matrices with integral entries, so the trace of $\hat\CL g$ must be integral as well. For the same reason, it is also clear that $(3)$ implies $(1)$. As for $(1)\Rightarrow (2)$, the idea of proof is as follows. First, one notices that $\Tr_V(\hat\CL g)\in \ZZ$ implies that $\Tr_V(\hat\CL \sum_i g_i)\in \ZZ$ for any collection $g_1,g_2,\ldots \in Co_0$. Then, using an explicit description of the lattice $\Lambda$ and of the generators of $Co_0$, one can show that for any two vectors $\lambda,\mu\in\Lambda$, there exists a sum $\sum_i g_i$ of elements in $Co_0$ such that
\be \Tr_V(\hat\CL \sum_i g_i)=\mu\cdot \hat\CL(\lambda)\ .
\ee If (1) holds,  then  $\mu\cdot \hat\CL(\lambda)$ must be integral for all $\lambda,\mu \in\Lambda$, and therefore $\hat\CL(\lambda)$ is in the dual lattice $\Lambda^*$. Using that the Leech lattice is self-dual, $\Lambda^*=\Lambda$, we conclude.

Finally, one needs to show that $(1)$ and $(2)$ imply $(3)$. To this aim, one identifies the $\RR$-linear maps $V\to V$ with vectors in $V\otimes V^*$ and the $\ZZ$-linear maps $\Lambda\to \Lambda$ with vectors in the (even self-dual) lattice $\Lambda\otimes\Lambda^*\subset V\otimes V^*$. We denote by $L\subseteq \Lambda\otimes \Lambda^*$ the sublattice generated by the $\ZZ$-linear maps $g:\Lambda\to \Lambda$ for all $g\in \Aut(\Lambda)$. In this notation, conditions $(1)$, $(2)$, and $(3)$ in theorem \ref{th:main} translate, respectively, to $\hat\CL\in L^*$, $\hat\CL \in \Lambda\otimes\Lambda^*$, and $\hat\CL\in L$. We already proved that $(1)\Leftrightarrow (2)$, so that $L^*=\Lambda\otimes  \Lambda^*$, and because $\Lambda\otimes \Lambda^*$ is self-dual, we conclude that $L=\Lambda\otimes  \Lambda^*=L^*$, i.e. all three conditions $(1)$, $(2)$, and $(3)$ in theorem \ref{th:main} are equivalent.

\bigskip

Theorem \ref{th:main} puts strong constraints on the fusion ring of the category $\calC_{SVir}(V^{f\natural})$. Let $K(\calC_{SVir}(V^{f\natural}))$ denote the Grothendieck ring of the fusion category $\calC_{SVir}(V^{f\natural})$, whose elements are formal finite integral linear combinations $\sum n_i[\CL_i]$, $n_i\in \ZZ$, $\sum |n_i|<\infty$, of isomorphism classes $[\CL_i]$ of simple objects in $\calC_{SVir}(V^{f\natural})$, and with product given by the fusion product of simple defects. An obvious consequence of theorem \ref{th:main} is:
\begin{corollary} There is a surjective (but not injective) ring homomorphism $K(\calC_{SVir}(V^{f\natural}))\to \End_\ZZ(\Lambda)$ from the Grothendieck ring of $\calC_{SVir}(V^{f\natural})$ to the ring of $\ZZ$-linear endomorphisms of the Leech lattice $\Lambda$.
\end{corollary}
Notice that, by theorem \ref{th:main}, even if we restrict to the subcategory of $\calC_{SVir}(V^{f\natural})$ generated by invertible defects $\CL_g$, $g\in Co_0$, the homomorphism is surjective and not injective.

\bigskip

There is a close (and mysterious) relation between the subgroups $H_\Pi\subset G\cong Co_0$ of $V^{f\natural}$-automorphisms fixing a $4$-dimensional subspace $\Pi\subset V_{tw}(1/2)$ of Ramond ground states, and the groups of symmetries of a family of non-holomorphic conformal field theories known as non-linear sigma models (NLSM) on K3.  The chiral and anti-chiral algebras of NLSM on K3 contain the $N=4$ superVirasoro algebra at central charges $c=\bar c=6$; the space of Ramond-Ramond ground states in these CFTs is always $24$-dimensional, all with positive fermion number. It was proved in \cite{Gaberdiel:2011fg} that if $H$ is a group of symmetries  of a NLSM on K3 preserving the (holomorphic and anti-holomorphic) $N=4$ superconformal algebras, as well as the spectral flow isomorphism, then $H$ is isomorphic to the subgroup $H_\Pi$ of $Co_0=\Aut(\Lambda)$ fixing a $4$-dimensional subspace $\Pi\subset \Lambda\otimes\RR$. Vice versa, every such subgroup $H_\Pi$ of $Co_0$ arises as a group of symmetries of some NLSM on K3. This correspondence between symmetries of $V^{f\natural}$ and symmetries of NLSM on K3 goes far beyond an isomorphism of abstract groups. It was noticed in \cite{Duncan:2015xoa} that for each such subgroup $H_\Pi\subset Co_0$, there is an (almost) perfect matching between the $g$-twining genera $\phi_g$ (a refined version of the McKay-Thompson series) computed in $V^{f\natural}$ and the ones obtained in the NLSM on K3. There is no known explanation for this coincidence.

There is a natural way to generalize this observation so as to include non-invertible topological defects. On one hand, one can consider the subcategory $\calC_{SVir,\Pi}(V^{f\natural})$ of $\calC_{SVir}(V^{f\natural})$ of topological defects $\CL$ that preserve a subspace $\Pi\subset V_{tw}(1/2)$. Equivalently, $\hat\CL$ acts on $\psi\in \Pi$ in the same way as on the vacuum, i.e. by multiplication by the quantum dimension $\langle \hat\CL\rangle$. This implies that the field $\psi(z)$ can be moved across the line defect $\CL$ without changing the correlation function. By theorem \ref{th:main}, there is a ring homomorphism from the Grothendieck ring of $\calC_{SVir,\Pi}(V^{f\natural})$ to a certain subring of $\End_\ZZ(\Lambda)$ that depends on $\Pi$. The elements of this subring are all the $\ZZ$-linear maps $L:\Lambda\to\Lambda$ such that, when extended to $\Lambda\otimes\RR$ by linearity, they preserve the orthogonal decomposition $\Lambda\otimes\RR=\Pi\oplus \Pi^\perp$ (i.e. $L(\Pi)\subseteq \Pi$ and $L(\Pi^\perp)\subseteq\Pi^\perp$),  and whose restriction to $\Pi\subset \Lambda\otimes \RR$ give maps $\Pi\to\Pi$ proportional to the identity.

Similarly, for NLSM on K3, one can consider the category $\cTop_\Pi$ of topological defects preserving both the holomorphic and anti-holomorphic $N=4$ superVirasoro algebras, as well as the spectral flow. It was shown in \cite{Angius:2024evd} that there is a homomorphism between the Grothendieck ring of $\cTop_\Pi$ and the same subring of $\End_\ZZ(\Lambda)$ considered above. While we have not been able yet to check whether the `defect twining genera' on the two sides match, these results suggest that the mysterious relation between symmetries of $V^{f\natural}$ and of NLSM on K3 extends to topological defects.

\bigskip

{\bf Acknowledgments.} I would like to thank all participants to the special session on `New Developments in infinite dimensional Lie algebras, vertex operator algebras and the Monster' at the UMI-AMS joint conference in Palermo, 25-26 July 2024, and in particular the co-organizers Darlayne Addabbo and Lisa Carbone. I am grateful to Scott Carnahan, Miranda Cheng, Chongying Dong,  John Duncan, Terry Gannon, Gerard H\"ohn, Theo Johnson-Freyd, Ching Hung Lam, Sven M\"oller, Niels Scheithauer for many useful conversations on the topics in this article. Some of the results presented here were obtained in collaboration with Roberta Angius, Stefano Giaccari and Sarah Harrison, whom I thank.
I acknowledge financial support from
the PRIN Project n. 2022ABPBEY, “Understanding quantum field theory through its deformations”, and from the CARIPARO Foundation Grant “Ricerca Scientifica di Eccellenza 2023” under project n. 68079 “New Tools for String Theory”.

 
\bibliographystyle{utphys}

\bibliography{mybib}


\end{document}